\documentclass[a4paper,11pt]{article}

\usepackage[utf8]{inputenc}
\usepackage[T1,T2A]{fontenc}
\usepackage{amsmath,amsfonts,amssymb}

 \usepackage{graphicx}
\usepackage{wrapfig}
\usepackage{subfig}
\usepackage{hyperref}
\usepackage[dvipsnames]{xcolor}

\begin{document}

\begin{titlepage}

\begin{center}
{\LARGE {\bf Gravitational production of sterile neutrinos}} \\
\vspace{2cm}
{\bf Fotis Koutroulis$^{\,1}$, Oleg Lebedev$^{\,2}$, Stefan Pokorski$^{\,1}$}
\end{center}

\begin{center}
  \vspace*{0.25cm}
  $^1$\it{Institute of Theoretical Physics, Faculty of Physics, University of Warsaw,\\
   ul. Pasteura 5, 02-093
Warsaw, Poland} \\
      \vspace{0.2cm}
 $^2$\it{Department of Physics and Helsinki Institute of Physics,\\
  Gustaf H\"allstr\"omin katu 2a, FI-00014 Helsinki, Finland}\\
\end{center}

\vspace{2.5cm}

\begin{center} {\bf Abstract} \end{center}
\noindent
We consider gravitational production of singlet fermions such as sterile neutrinos during and after inflation.
The production efficiency due to classical gravity is suppressed by the fermion mass.
Quantum gravitational effects, on the other hand,  
are expected to break conformal invariance of the fermion sector by the Planck scale--suppressed operators  
irrespective of the mass. We find that such operators are very efficient in fermion production immediately after inflation,
generating a significant background of stable or long-lived feebly interacting particles. This applies, in particular, to sterile neutrinos
which can constitute cold non--thermal dark matter for a wide range of masses, including the keV scale.

\end{titlepage}


\tableofcontents

\vspace{1cm}

\section{Introduction}

The existence of right-handed neutrinos $\nu_R$ is motivated by the small but non-zero masses of the active neutrinos
\cite{Minkowski:1977sc,GellMann:1980vs,Yanagida:1979as,Mohapatra:1979ia,Schechter:1980gr,Lazarides:1980nt}. In addition to generating masses, $\nu_R$ can be relevant to the problem of dark matter
(DM).
Indeed, light (mostly) right-handed neutrinos, which we will also call ``sterile'' neutrinos, can have a lifetime longer than the age of the Universe and also have the properties characteristic of dark matter, e.g.
very weak  interactions with the Standard Model (SM) states.
This makes $\nu_R$ an attractive minimalistic dark matter candidate \cite{Dodelson:1993je,Shi:1998km,Abazajian:2001nj}, as reviewed in \cite{Boyarsky:2009ix,Boyarsky:2018tvu}.

    \begin{figure}[h!] 
\centering{
\includegraphics[scale=0.45]{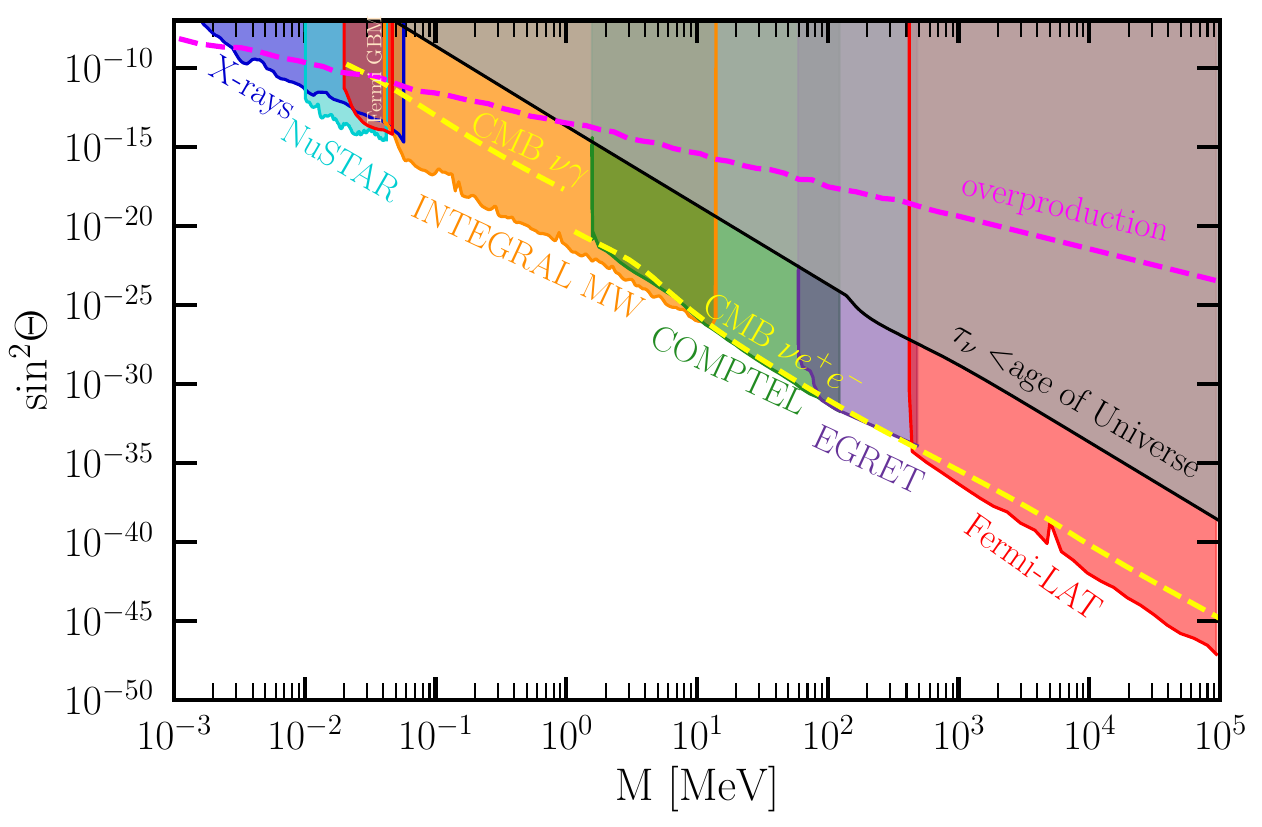}
}
\caption{ \label{par-space} {\small
Constraints on the active-sterile neutrino mixing angle $\Theta$ from astrophysics and cosmology.  The figure is from Ref.\,\cite{Lebedev:2023uzp}.
 }}
\end{figure}

It is natural to assume that there are 3 right-handed neutrinos, although there could be many more of them \cite{Buchmuller:2007zd}. The two heavier $\nu_R$ would then be responsible for the active neutrino masses, while the lightest one can play the role of dark matter \cite{Asaka:2005an,Asaka:2006nq}. This is possible if the sterile-active mixing angle is tiny, which makes the lightest $\nu_R$ long lived.
The cosmological and astrophysical  constraints on this angle are shown in Fig.\,\ref{par-space} (see \cite{DeRomeri:2020wng} for further details). The most important processes are the decays 
$\nu_R \rightarrow \nu_a \gamma$ and $\nu_R \rightarrow \nu_a e^- e^+$, where $\nu_a$ is the active neutrino. These lead to the X-ray and gamma ray emission as well as to the CMB distortion,
which set significant constraints on decaying neutrinos.

If the mixing angle is not too small, sterile neutrinos are produced by the Standard Model thermal bath via the Dodelson-Widrow mechanism 
\cite{Dodelson:1993je}. 
In principle, this could generate  the right amount of ``warm'' dark matter, although this possibility is 
now disfavored \cite{Boyarsky:2005us,Seljak:2006qw,Boyarsky:2006fg,Yuksel:2007xh,Perez:2016tcq,Ackermann:2015lka,Adhikari:2016bei}. The corresponding constraint is shown in the figure
by the ``overproduction'' line. In particular, substantial mixing angles are ruled by overabundance of dark matter.

The thermal production mechanism assumes that the initial abundance of sterile neutrinos is zero. We show that this assumption is not quite realistic  \cite{Lebedev:2022cic}
since particles are produced during and after inflation via gravitational effects.
In particular, Planck-suppressed operators induced by quantum gravity 
 play an important role during the inflaton oscillation epoch  \cite{Lebedev:2022ljz}  and can readily dominate production of light fermions.
The characteristic particle energy is far below the Standard Model bath temperature which makes such fermions good cold dark matter candidates, in contrast to the particles produced via the thermal 
emission.

In general, gravitational particle production creates a significant background of dark relics, which affects the predictions of most non-thermal dark matter models \cite{Lebedev:2022cic}. 
 Hence, predictive models require either excellent  control over quantum-gravity induced operators or a mechanism for dilution of gravitationally produced particles. An example of the latter is 
 provided by models with an extended period of matter domination resulting in a low reheating temperature \cite{Cosme:2023xpa}.

In what follows, we study gravitational fermion production during inflation and in the inflaton oscillation epoch. In these periods, the energy density and the field values are the largest, leading to   most 
efficient particle production.

\section{Fermion production during inflation }

Generally, particles are produced due to the expansion of the Universe \cite{Parker:1969au,Grib:1976pw,Chung:1998zb}, which  can be attributed to the change in the vacuum state in a time-dependent  background (see \cite{Ford:2021syk} for a review).
In what follows, we study in detail fermion production in the Friedmann Universe \cite{Parker:1971pt}. 
For our exposition to be self-contained,  we start  with a pedagogical  introduction following  Ref.\,\cite{Chung:2011ck}.

Consider a Dirac fermion $\Psi$ of mass $M$ which has negligible couplings to other fields. The corresponding results for a Majorana fermion can be obtained by a simple rescaling. We assume the fermion to be light enough relative to the 
Hubble rate during inflation, $M \ll H$, so that there is no ``energetic'' obstacle to its production.
 Since the Friedmann metric is conformally flat and the fermion action is conformally invariant apart from the mass term, particle production  via classical gravity  is fully controlled by the fermion mass $M$.
 In what follows, we verify this explicitly and compute the resulting abundance of $\Psi$.
 
 \subsection{Basics}
 
 The starting point is the Dirac equation in curved space,
\begin{equation}
(i \gamma^\alpha \nabla_\alpha- M) \,\Psi =0 \;,
\end{equation}
which follows from the action $\int d^4 x \sqrt{|g|} \bar \Psi (i \gamma^\alpha \nabla_\alpha- M) \,\Psi $, where $g_{\mu\nu}$ is the space-time metric, $\nabla$ is the covariant derivative and $\alpha$ is the  local Lorentz index.
The Friedmann metric in terms of the conformal time $x_0 \equiv \eta$ reads
\begin{equation}
ds^2 = a(x_0)^2 \, \eta_{\mu \nu } dx^\mu dx^\nu \;.
\end{equation}
Using the Weyl transformation
\begin{equation}
g_{\mu \nu }  = \Omega^2 \tilde g_{\mu \nu } ~,~ \Psi = \Omega^{-3/2} \tilde \Psi ~,~e_\alpha^\mu = \Omega^{-1} \tilde e_\alpha^\mu ~,
\end{equation}
where   $\Omega= a(x_0)$ and   $e_\alpha^\mu$ is the vierbein, 
the factor $a(x_0)$ can be eliminated from the action apart from the mass term.\footnote{This requires conservation of the vector current, $\nabla_\mu\; \bar \Psi \gamma^\mu \Psi$=0.} 
 Dropping the tilde over the transformed quantities, the resulting Dirac equation reads
\begin{equation}
(i \gamma^\mu \partial_\mu-  a(\eta) M) \,\Psi =0 \;,
\end{equation}
which is the flat space Dirac equation with a time-dependent mass. The latter causes 
 particle production.

 The above equation can be solved as follows. 
 The solution space is spanned by the orthonormal basis $\{  U,V \}$, where 
 the basis vectors  characterized by the 3-momentum ${\bf k}$ and the spin projection  $s$  have the form
  \begin{equation}
 U_{{\bf k} ,s} (\eta, {\bf x})= {e^{i {\bf k} \cdot {\bf x}}    \over (2 \pi)^{3/2}} 
 \left(
 \begin{matrix}
  u_{A,k} (\eta) \\
  s \, u_{B,k} (\eta)
 \end{matrix}
 \right)   \otimes h_s ({\bf \hat k}) \;,
 \label{U}
 \end{equation}
 where $k \equiv |{\bf k}|$, ${\bf \hat k} = {\bf k } / |{\bf k}|$, $h_s$ are the helicity 2-spinors satisfying
 \begin{equation}
 {\bf \hat k} \cdot \vec{ \sigma} \; h_s = s \, h_s ~~,~~ s= \pm 1 \;,
 \end{equation}
 and $\vec{\sigma}$ are the sigma matrices. $u_{A,B}$ are complex functions of time  to be determined, depending on $a(\eta)$.
In spherical coordinates, ${\bf \hat k}  = (\theta, \phi)$ and 
 \begin{equation}
 {\bf \hat k} \cdot \vec{ \sigma}  = 
 \left(
 \begin{matrix}
 \cos \theta & e^{-i \phi} \, \sin\theta  \\
   e^{i \phi} \, \sin\theta & -\cos\theta
 \end{matrix}
 \right) ~,~ 
 h_{-1} =  \left(
 \begin{matrix}
  e^{-i \phi} \, \sin {\theta\over 2} \\
   -\cos {\theta\over 2}
 \end{matrix}
 \right)~,~
 h_{1} =  \left(
 \begin{matrix}
  e^{-i \phi} \, \cos {\theta\over 2} \\
   \sin {\theta\over 2}
 \end{matrix}
 \right)\;.
  \end {equation}
 In this convention, $ -i \sigma^2 h_s^*({\bf \hat k} )= -s e^{i \phi} h_{-s}({\bf \hat k} )$ and $h_s(-{\bf \hat k} )  =- h_{-s}({\bf \hat k} )$. 
 
 The $V$-vectors can be chosen as $V_i= -i \gamma^2 U_i^*$, so that\footnote{Note the (inconsequential) phase difference from the result in \cite{Chung:2011ck}. }
 \begin{equation}
 V_{{\bf k} ,s} (\eta, {\bf x})= -{e^{-i {\bf k} \cdot {\bf x}}    \over (2 \pi)^{3/2}} 
 \left(
 \begin{matrix}
  -u_{B,k}^* (\eta) \\
  s \, u_{A,k}^* (\eta)
 \end{matrix}
 \right) \otimes  h_s (-{\bf \hat k})  \, e^{i \phi}\;,
 \end{equation}
 in the convention
 \begin{equation}
 \gamma^0 = 
 \left(
 \begin{matrix}
I & 0 \\
 0 & -I
 \end{matrix}
 \right) ~,~ 
 \gamma^i =  \left(
 \begin{matrix}
0& \sigma^i \\
 -\sigma^i & 0
 \end{matrix}
 \right) ~.
 \end{equation}
 Since 
\begin{equation}
h^{\dagger}_s ({\bf \hat k} ) \,h_r ({\bf \hat k} ) = \delta_{rs} \;,
\end{equation} 
 the orthonormality of the basis,
 \begin{equation}
 (U_i, U_j)= \delta_{ij}~, ~  (U_i, V_j)= 0\;,
 \end{equation}
 requires 
  \begin{equation}
|u_A|^2 + |u_B|^2 =1\;.
\label{norm}
 \end{equation}
Here the scalar product is meant in the usual sense, $(f,g) = \int d^3 x \, f^\dagger g$, and following  \cite{Ford:2021syk}, we take index $i$ to be continuous. 
In particular, the spacial part of the wave functions is described by the orthonormal set  $ {e^{i {\bf k} \cdot {\bf x}}    \over (2 \pi)^{3/2}} $.

  Using the above Ansatz, the  equation of motion (EOM) reduces to 
 \begin{equation}
 i\partial_\eta        
 \left(
 \begin{matrix}
  u_A \\
  u_B
 \end{matrix}
 \right)
  = \left(
 \begin{matrix}
aM & k \\
 k & -aM
 \end{matrix}
 \right) \; 
 \left(
 \begin{matrix}
  u_A \\
  u_B
 \end{matrix}
 \right)
 ~.~ 
 \label{EOM-u}
 \end{equation}
  This implies, in particular, that the $u_{A,B}$ normalization  (\ref{norm}) is time-independent.  Note that the evolution in the $(u_A, u_B)$ space is unitary, i.e. SU(2). 
  Denoting the time derivative  $\partial_\eta$ by a prime, we reduce the system  to second order differential equations,
  \begin{eqnarray}
  && u_A^{\prime \prime} + (iM a^\prime +a^2 M^2 +k^2) \,u_A=0 \,\\
  && u_B^{\prime \prime} + (-iM a^\prime +a^2 M^2 +k^2) \,u_B=0 \,,
  \label{uA-eq}
  \end{eqnarray}
  where $k$ is the magnitude of the 3-momentum.
 The solutions must have certain asymptotic behaviour corresponding to the $in$ or $out$ vacuum. In particular, during inflation the solutions are Hankel functions of $\eta$.
 
 In the Heisenberg picture, the field operator is expressed via creation/annihilation operators times the basis functions solving the Dirac equation,
 \begin{equation}
 \Psi (x) = \sum_i  \left( a_i U_i + b_i^\dagger V_i \right) \;.
 \end{equation}
 Here the operators satisfy the usual time-independent anti-commutation relations, $\{  a_i , a^\dagger_j\}= \delta_{ij}$, $\{  b_i , b^\dagger_j\}= \delta_{ij}$, etc.
The vacuum is defined by $a_i | 0 \rangle = b_i | 0 \rangle =0$. The Hilbert space is constructed  via Fock states  by acting with creation operators on the vacuum.

 The creation/annihilation operators are attached to a specific solution basis. Since the basis is complete at a given $\eta$, a new set of basis functions can be expressed as
\begin{equation}
\tilde U_i = \sum_j \left(   \alpha_{ij} U_j + \beta_{ij} V_j \right)~~,~~
\tilde V_i = \sum_j \left(   \alpha_{ij}^* V_j + \beta_{ij}^* U_j \right) \;.
\end{equation}
where the second relation follows from the first one.
Using orthonormality of the basis, one has
\begin{equation}
\beta_{ij}=(V_j , \tilde U_i) ~~, ~~ \alpha_{ij}=(U_j , \tilde U_i) \;.
\label{UV}
\end{equation}
The new basis is also orthonormal, which together with 
\begin{equation}
\Psi =\sum_i  \left( a_i U_i + b_i^\dagger V_i \right) = \sum_i  \left( \tilde a_i \tilde U_i + \tilde b_i^\dagger \tilde V_i \right) \;,
\end{equation}
implies
\begin{equation}
\tilde a_i = \sum_j \left( \alpha^*_{ij} a_j +   \beta^*_{ij} b^\dagger_j   \right)\;, 
\end{equation}
such that
\begin{equation}
 \langle \tilde N_i \rangle \equiv  \langle 0 | \tilde a_i^\dagger \tilde a_i | 0 \rangle = \sum_j | \beta_{ij}|^2  \;.
 \label{N}
\end{equation}
This gives the mean number of $tilded$ particles of  type $i$ in the original vacuum $|0\rangle$ defined by the absence of any un-tilded particles.

 Now let us consider a particular set of basis transformations which affects $u_{A,B}$ while leaving  the spin and spacial  components of the wave functions intact, i.e. preserves the Ansatz (\ref{U}).
 Since $\{ U,V \}$ is a complete basis,   the $U$-vectors transform as
\begin{equation}
\tilde U_{{\bf k} ,s} = \alpha_{{\bf k} ,s} U_{{\bf k} ,s} + \beta_{{\bf k} ,s} V_{-{\bf k} ,s} 
\end{equation}
under the basis change by virtue of (\ref{UV}). 
Here we use a shorthand notation for the indices of $\alpha$ and $\beta$: $ij \rightarrow {{\bf k} ,s}$ since there is only one term in the sum.
Employing  the explicit parametrization of $\tilde U, \tilde V$ as in (\ref{U}) and computing various scalar products, one
finds
\begin{equation}
\beta_{{\bf k} ,s} = {\rm phase } \times  (u_{A,k} \tilde u_{B,k} -u_{B,k} \tilde u_{A,k}) \;,
\end{equation}
where the (time-independent)  phase factor is irrelevant for our purposes.
Here we have used the fact  $u_{A,B}$ depend only on the magnitude of the 3-momentum  as is clear from (\ref{uA-eq}).

The EOM (\ref{EOM-u}) imply that   $\beta$ is time-independent,
\begin{equation}
\beta^\prime =0 \; ,
\end{equation}
which can also be viewed as conservation of the cross product of 2 vectors under SU(2) rotations.
This conservation law is important since it allows for evaluation of $\beta$ at any convenient point in time.

 \subsection{Particle number calculation}
 
 The number of particles produced by inflation is given by   $|\beta|^2$  for the 
 2 bases corresponding to the $in$ and $out$ states, respectively. 
 The physical picture is as  follows: we define the system initially in the $in$ state ($\eta \rightarrow -\infty$) with no particles, 
 while the observed particle number is measured with respect to the  $out$ vacuum 
($\eta \rightarrow \infty$),
 as  given by (\ref{N}) in the Heisenberg picture.
  Eq.\,\ref{EOM-u}  implies that,   far in the past,  the system is Minkowskian since $a\rightarrow 0$ and only the $k$ terms matter. In the future, it is also Minkowskian: $a\rightarrow \infty$ but
 $H \rightarrow 0$, which makes the system  effectively static, while the momentum terms can be neglected. 
 Hence the $in$ and $out$ vacua are those of flat space. As usual, the mode functions are the  positive frequency $\omega$ modes, $e^{-i\omega \eta}$, or more precisely,
 \begin{equation}
 \left(
 \begin{matrix}
  u_A \\
  u_B
 \end{matrix}
 \right) \propto e^{-i \int \omega (\eta) d \eta}
 \end{equation}
 with $\omega >0$. This fixes the $in$ and $out$ boundary conditions in the asymptotic regions.
 
 Naturally, the result depends on $a(\eta)$, in particular, whether inflation is followed by a radiation-dominated or a matter-dominated expansion period. We consider both possibilities in what follows. 

\subsubsection{Inflation followed by radiation domination}

The function $a(\eta)$ is chosen such that it describes a smooth transition from inflation at early times to radiation domination at late times \cite{Chung:2011ck},
 \begin{eqnarray}
 && a(\eta) = \left\{  
 \left(  {1\over a_e H_e}  -\eta  \right)^{-1} H_e^{-1}
  ~~{\rm  for }~~ \eta\leq 0 ~~,~~ a^2_e H_e \left( \eta + {1\over a_e H_e}\right) ~ ~{\rm for}~ \eta>0 \right\}~,\\
 && H(\eta) = \left\{H_e ~~{\rm for}~~ \eta \leq 0 ~~,~~ H_e \, (a_e/a)^2 ~ ~{\rm for}~ ~\eta>0 \right\}~,
 \end{eqnarray}
 where $a_e$ and $H_e$ are the scale factor and the Hubble rate at the end of inflation, respectively.
 Note that $H(\eta)= a^\prime /a^2$ in terms of the conformal time.

 At $\eta \rightarrow - \infty$, the $aM$ terms in (\ref{EOM-u}) can be neglected and the positive  eigenstate of the matrix on the right hand side (RHS) is $(1/\sqrt{2}, 1/\sqrt{2})^T$. Hence the positive frequency solution is
  \begin{equation}
 \left(
 \begin{matrix}
  u_A \\
  u_B
 \end{matrix}
 \right) \overset{\small{\eta \rightarrow -\infty}}{\longrightarrow} 
 \left(
 \begin{matrix}
  {1/ \sqrt{2}} \\
 {1/ \sqrt{2}}
 \end{matrix}
 \right) \;e^{-i k \eta} \;.
 \end{equation}
 This fixes uniquely the $in$ solution in the inflationary regime.
 
 At $\eta \rightarrow  \infty$, the matrix is diagonal and the positive eigenvalue solution corresponds to
  \begin{equation}
 \left(
 \begin{matrix}
  u_A \\
  u_B
 \end{matrix}
 \right) \overset{\small{\eta \rightarrow \infty}}{\longrightarrow} 
 \left(
 \begin{matrix}
 1\\
0
 \end{matrix}
 \right) \;e^{-i \int \omega(\eta) d\eta} \;,
 \label{eta-inf}
 \end{equation}
with $\omega \rightarrow a(\eta) M$. This fixes uniquely the $out$ solution in the radiation domination regime.

In what follows, we construct analytical solutions for $u_A, u_B$ away from $\eta \sim 0$.
In this case, 
$a(\eta)$ can be approximated by $- {1\over H_e \eta}$ during inflation and by $a^2_e H_e \eta$ during radiation domination.

{\bf \underline{Inflationary regime.}}
 In the inflationary regime, $u_A$ satisfies
 \begin{equation}
 \eta^2 u_A^{\prime \prime} + \left(    k^2 \eta^2 +     \left[   {iM \over H_e} + {M^2 \over H_e^2}  \right]  \right)\, u_A =0 \;.
 \end{equation}
 This is the Bessel-type equation\footnote{
 The equation $x^2 y^{\prime\prime} +(2p+1) xy^{\prime} +(a^2 x^{2r}+ \beta^2)y=0$ is solved by
 $y=x^{-p}\left[     C_1 \,J_{q/r} (\alpha x^r/r)   + C_2 \,Y_{q/r} (\alpha x^r/r)   \right]$.}, whose solution with the right asymptotics are the Hankel functions. One  finds
  \begin{equation}
u_A (\eta) = \sqrt{\pi k \eta \over 4}\,  e^{- i {\pi \over 2} (1-iM/H_e)}\, H^{(2)}_{{1/ 2} -{iM/H_e}} (k\eta) \;,
 \end{equation}
 assuming $M /H_e \ll 1$.
 Using $H^{(2)}_\nu (-z) = - e^{i\nu \pi} H_\nu^{(1)} (z)$ and $\sqrt{-1} =-i$ according to the phase convention of the Hankel argument, one can rewrite it as
 \begin{equation}
u_A^{in} (a) =  \sqrt{\pi k  \over 4 a H_e}\, 
  e^{ i {\pi \over 2} (1-iM/H_e)}\, H^{(1)}_{{1/ 2} -{iM/H_e}} \left({k\over a H_e}\right) \;.
 \end{equation}
 The $u_B$ function is obtained by flipping the sign of $M$,
 \begin{equation}
u_B^{in} (a) = \sqrt{\pi k  \over 4 a H_e}\, 
  e^{ i {\pi \over 2} (1+iM/H_e)}\, H^{(1)}_{{1/ 2} +{iM/H_e}} \left({k\over a H_e}\right) \;.
 \end{equation}

 {\bf \underline{Radiation-dominated regime.}}
 The equation for $u_A$ is
 \begin{equation}
  u_A^{\prime \prime} + \left(    k^2 + iM a_e^2 H_e + \eta^2 \, M^2 a_e^4 H_e^2  \right)\, u_A =0 \;.
  \label{uA-rad}
 \end{equation}
The solution is a parabolic cylinder function $D_\nu (z)$.\footnote{The equation $y^{\prime\prime} +(\nu +1/2 -z^2/4) y=0$ is solved by $D_\nu(z)$, $D_{-\nu-1} (iz)$.}
Defining 
\begin{equation}
  C={k^2 \over 2M a_e^2 H_e} \;,
 \end{equation}
we find
\begin{equation}
  u_A^{out} (\eta) = e^{-{\pi\over 4} C} D_{-iC} \left( e^{i\pi/4 } \sqrt{2M \over H(\eta)} \right) \times {\rm phase} \;,
 \end{equation}
where the  time-dependent  phase is universal for $u_A$ and $u_B$, and thus irrelevant for our purposes.\footnote{This phase is suppressed by $\ln \eta/ \eta^2$ and thus vanishes at large $\eta$.}

 The equation for $u_B$ is obtained by $M \rightarrow -M $ in (\ref{uA-rad}). The solution vanishing at infinity is 
 \begin{equation}
  u_B^{out} (\eta) = \sqrt{C } \,e^{-{\pi\over 4} C + {i\pi \over 4}} D_{-1-iC} \left( e^{i\pi/4 } \sqrt{2M \over H(\eta)} \right) \times {\rm phase} \;,
  \label{uB-rad}
 \end{equation}
 where the  ``phase'' is the same as that in $u_A$. 
 The solution  approaches zero as $1/ \eta$.

We note that (\ref{eta-inf}) does not fix the normalization of $u_B$ since it vanishes at infinity. 
To determine the normalization factor, 
one needs to make the asymptotic behaviour of the positive frequency mode more precise by including the $k/aM$ correction to the  eigenstate of the matrix in  (\ref{EOM-u}),
 \begin{equation}
 \left(
 \begin{matrix}
  u_A \\
  u_B
 \end{matrix}
 \right) \overset{\small{\eta \rightarrow \infty}}{\longrightarrow} 
 \left(
 \begin{matrix}
 1\\
{k\over 2aM}
 \end{matrix}
 \right) \;e^{-i \int \omega(\eta) d\eta} \;,
 \label{eta-inf1}
 \end{equation}
 while the correction to $\omega$ can be neglected. This shows that $u_B$ vanishes as $1/\eta$ and also fixes the normalization as in (\ref{uB-rad}).
 
{\bf \underline{Particle production.}}
  We have obtained the   solutions which are valid in two regimes: the $in $ solution works during inflation and the $out$ solution works only in the radiation-dominated regime.
  At the end of inflation  $a\sim a_e$, corresponding to the transition region,  both of them are approximately valid. Since $\beta$ is time-independent for the exact solutions, we can compute  
   $\beta$  at this point.  The validity of this approximation is supported by numerical analysis.
    
    The average particle number with momentum $k$ is computed via
 \begin{equation}
|\beta_{{\bf k} ,s}|^2 = |u_{A,k}^{in} u_{B,k}^{out} -u_{B,k}^{in}  u_{A,k}^{out}|^2 \;,
\end{equation}
 and the total particle number density is \cite{Ford:2021syk}
 \begin{equation}
 n = \int {d^3 {\bf k} \over (2\pi)^3 a^3} \, |\beta_{{\bf k},s }|^2 \;.
 \label{n}
 \end{equation}
 Given the analytical results, one can compute $\beta_k (a_e)$ numerically for different $k$. One finds that  for $M\ll H_e$, 
 the $in$ solutions at $a\sim a_e$  approximately retain their asymptotic form, while the $out$ solutions can change drastically. 
 The main factor determining  $\beta_k (a_e)$ is the value of $C$:
 \begin{eqnarray}
 && C \ll 1 \Rightarrow |\beta_k|^2  \sim 1/2 \;, \\ 
 &&  C \gg 1 \Rightarrow |\beta_k|^2  \sim 0 \;,
 \end{eqnarray}
 so $C \sim 1$ corresponds to the cut-off of particle production and 3-momenta  above  $ a_e \sqrt{H_e M} $ are not generated. The  $|\beta_k|^2$ fall-off with $k$ is fast, $\propto k^{-6} $ at large $k$,
 hence these modes do not contribute to the integral in any significant way.  
 
 This qualitative behavior can be understood analytically.  The   $in$ state does  not change significantly from $a\sim 0$ to $a_e$ independently of the momenta as long as $M \ll H_e$,
 so 
 \begin{equation}
 \left(
 \begin{matrix}
  u_A \\
  u_B
 \end{matrix}
 \right)^{in} (a_e) \sim 
 \left(
 \begin{matrix}
 1/\sqrt{2}\\
1/\sqrt{2}
 \end{matrix}
 \right) \times {\rm phase}\;,
 \end{equation}
 where the overall phase is irrelevant for our purposes.
For the out state, the zero argument limit of the $D$-functions, i.e. $H_e \gg M$, gives
  \begin{equation}
 \left(
 \begin{matrix}
  u_A \\
  u_B
 \end{matrix}
 \right)^{out} (a_e) \sim 
 \left(
 \begin{matrix}
 e^{-\pi C/4} \sqrt{\cosh {\pi C \over 2}}\\
 e^{-\pi C/4} \sqrt{\sinh {\pi C \over 2}}
 \end{matrix}
 \right) \times {\rm phase}\;.
 \end{equation}
 This explains why $C$ acts as the main driver of particle production efficiency: for $C\gg 1$, the $in$ and $out$ states coincide up to the phase and $\beta \sim 0$. For small $C$, only $u_A$ is significant for the $out$ state, so $\beta \sim 1/\sqrt{2}$.

 {\bf \underline{Relic abundance.}}
 We observe that the momentum cutoff corresponds to $C\sim 1$ so that production of particles with momenta larger than \cite{Chung:2011ck}
 \begin{equation}
 k_* =   a_e \sqrt{H_e M} 
 \end{equation}
 is suppressed.
 To understand the physics of this cut-off,  define $a_M$ as the moment when the Hubble rate equals the particle mass,
 \begin{equation}
      a_M :~ ~H(a_M)=M \;. 
 \end{equation}     
Since $H = H_e (a_e/a)^2$,
 \begin{equation}
 k_* = a_M \, M \;.
 \end{equation}
Therefore, 
{\it only  particles with physical 3-momenta  $k/a < M$ at the time $H=M$ are created.}
 In other words, these particles are {\it non-relativistic} or, at best, semi-relativistic. 
 
 Inserting the step function $\theta(k_* -k)$ in the integral (\ref{n}), we get \cite{Chung:2011ck}
 \begin{equation}
 n(a) \sim 4\times {1\over 2\pi^2 } \times {1\over 2} \, {M^3  \over 3} \, \left(   {a_M \over a } \right)^3 \;,
 \label{na}
 \end{equation}
 where 4 comes from the d.o.f. of the Dirac field and $1/2$ from $|\beta_k|^2$ of the excited momentum modes.
 This formula implies that particle production stops when $H=M$ and after that the total particle number is conserved. 
 The density is proportional to the conformal symmetry breaking parameter $M$ and  consistent with thermal interpretation of the de Sitter space:
 indeed $n \sim M^3$ is expected at $T\sim H_M =M$.

 The above result conforms to our expectations:
  massless particles (or highly relativistic ones) are not produced at all since the system becomes  scale-invariant in this case. 
 Note also that production of superheavy particles,  $M > H_{\rm infl}$,             would be suppressed, although this is 
 not immediately clear in our approximation  $M \ll H_e$ in the Hankel functions.

 The abundance $Y$ of created particles can be estimated at $a=a_M$, after which it remains constant. 
 It is defined by
  \begin{equation}
 Y= {n \over s_{\rm SM}} ~~,~~ s_{\rm SM} = {2\pi^2 \over 45}\, g_{*} T^3 \;,
 \label{Y-def}
  \end{equation}
where $ s_{\rm SM}$ is the entropy density of the SM thermal bath at temperature $T$ 
     and $g_{*}$ is the effective number of degrees of freedom contributing to the entropy.

Radiation domination after inflation can correspond either to inflaton oscillations in a $\phi^4$ potential  followed by reheating or instant reheating in an arbitrary potential,
both of which lead to the same scaling $a(\eta)$ and the same relic abundance of $\Psi$. 
 The reheating temperature $T_R$ is found via 
  \begin{equation}
  3 H_{ R}^2 M_{\rm Pl}^2 = {g_* (T_{R}) \pi^2 \over 30} \, T_{R}^4 \;,
  \end{equation}
 and $H_e/H_R = (a_R/a_e)^2$.
  Since $a_R/a_M = (M/H_R)^{1/2}$ and $s_{\rm SM} (a_M) = (a_R/a_M)^3 \, s_{\rm SM} (a_R)$,
 we have
 \begin{equation}
 Y\simeq 4.5 \times 10^{-3} \; \left({M \over M_{\rm Pl}}\right)^{3/2} \;,
 \label{Yrad}
 \end{equation}
 with no dependence on the Hubble rate (!) nor reheating temperature as long as $M\ll H_e$.
 The observational constraint on dark matter 
 \begin{equation}
 Y \leq 4.4 \times 10^{-10} \; {{\rm GeV} \over M } \;
  \label{Y0}
  \end{equation}
 then requires
\begin{equation}
M \lesssim 2 \times 10^8 \, {\rm GeV} \;,
\label{Mbound}
\end{equation}
 if the fermion $\Psi$ is $stable$ or very long-lived. This is independent of the Hubble rate during inflation as long as it is larger than the fermion mass,
 which we find quite remarkable.
 The above result implies that the abundance of lighter fermions ($M \ll 10^8$ GeV) is negligible and there are no useful constraints.

  Here we assume that $\Psi$ is Dirac,
 while for the Majorana fermion the abundance should be divided by 2 to account for two Majorana d.o.f.
 
 One may imagine  that the  fermion with  mass     $2 \times 10^8 \, {\rm GeV}$ 
 produced by gravity during inflation
  constitutes {\it all of the dark matter}. However,
 its density perturbations are not correlated with that of the inflaton, hence it is disfavored by isocurvature constraints.

 \subsubsection{Inflation followed by matter domination}
 
 It is possible that inflation is followed by a long period of matter domination. This is the case when the inflaton oscillations occur in a $\phi^2$ potential and the inflaton decays very slowly.
Particle production takes place in that period so that  the
 $out$ boundary conditions on the wave function should  be imposed during the matter domination era.  Although this possibility appears to be less common in the literature, we find it equally 
 viable\footnote{This happens, for example, when a heavy inflaton couples very weakly to the Higgs field leading to a low reheating temperature
 (see, e.g. \cite{Lebedev:2021xey}).}.

 The Hubble rate scales as $a^{-3/2}$, therefore   solving $a^\prime/a^2 = H_e (a_e/a)^{3/2}$ with the boundary condition at $a_e$ corresponding to $\eta=0$, we get
 \begin{equation}
 H=H_e \left( {a_e\over a}\right)^{3/2} ~~,~~ a= {1\over 4} a_e^3 H_e^2 \left( \eta + {2\over a_e H_e}   \right)^2 \;.
 \end{equation}
 At $\eta \gg 1/(a_e H_e)$, the EOM for $u_A$ reads
 \begin{equation}
 u_A^{\prime \prime} + \left(     {i\over 2}  M H_e^2 a_e^3\,\eta   +  {1\over 16} M^2 H_e^4 a_e^6\, \eta^4 +k^2   \right)\, u_A =0 \;,
 \end{equation}
 while the EOM for $u_B$ is obtained by the replacement $M \rightarrow -M$.
 The oscillation frequency squared is now a quartic polynomial in time and the exact solution is challenging to find.
 Hence, we solve the equation numerically.
 The oscillation frequency at late times is $\omega \simeq a M = {1\over 4} M H_e^2 a_e^3 \, \eta^2$ such that the boundary condition at $\eta \rightarrow \infty$ becomes
 \begin{equation}
 \left(
 \begin{matrix}
  u_A \\
  u_B
 \end{matrix}
 \right)^{out}  \simeq 
 \left(
 \begin{matrix}
       1 \\
  {2k\over  M H_e^2 a_e^3 \, \eta^2} 
 \end{matrix}
 \right) \times    e^{ -{i\over 12 } M H_e^2 a_e^3 \, \eta^3}    \;.
 \end{equation}
Using the  inflationary $in$ states as before, we then compute $\beta_k^2$   at $\eta\sim 0$ finding that the effective momentum  cut-off for particle production is
  \begin{equation}
k_* \sim M^{1/3} H_e^{2/3} a_e \;,
 \end{equation}
 corresponding to $k_* = a_M M$ as before. 
 
 This result can be understood qualitatively from the EOM.
 As one goes from large $\eta$ to its smaller values for negligible  $k \rightarrow  0$,  
 the vector $(1,0)^T$ remains an eigenvector of the frequency matrix.
 So, $u_B$ stays close to zero and $\beta_k$ takes on its near-maximal value $\sim 1/\sqrt{2}$. 
   As one increases $k$, the $(u_A, u_B)^T $ vector starts to rotate. 
   The $k^2$ term becomes significant  in the EOM when $M^2 H_e^4 a_e^6\, \eta^4 \sim k^2$. If the damping term is also substantial at this time,
 the magnitude of $u_A$ decreases and that of $u_B$ increases due to the constraint $|u_A|^2 + |u_B|^2=1$. So, the transition to a different regime occurs when all the three terms in $\omega^2$ become equally important,
 \begin{equation}
 M H_e^2 a_e^3\,\eta \sim M^2 H_e^4 a_e^6\, \eta^4 \sim k^2 \;,
 \end{equation}
 which gives the above value of $k_*$.
 
 {\underline {\bf Relic abundance}}
 The result (\ref{na}) applies also in the matter dominated case, although the expression for $a_M$ changes. It implies that particle production stops at $H \sim M$, as before.
 Our matter domination assumption means that 
 reheating occurs after particle production terminates,
 \begin{equation}
 H_M = M > H_R \;. 
 \end{equation}
 The relic abundance is computed at reheating, after which it remains constant. 
 $n(a_R)$ is obtained 
   from  (\ref{na})  using  the scaling
 $ 
 \left(  {a_M \over a_R} \right)^3 = {\rm const} \times {T_R^4 \over M^2 M_{\rm Pl}^2} \;,
 $
 where the const depends on the number of d.o.f.
 One then finds for a Dirac fermion,
 \begin{equation}
 Y = 8 \times 10^{-3} \;{M \,T_R \over M_{\rm Pl}^2} \;,
 \end{equation}
where  $g_* \sim 107$ has been assumed.
 This is smaller than the radiation dominated result (\ref{Yrad}) due to the constraint $M > H_R$, which implies $(M/M_{\rm Pl})^{1/2} > 1.8 \, T_R /M_{\rm Pl}$. For a Majorana fermion,
 the above $Y$ is to be divided by two.
 
 Requiring the abundance of the fermion to be below that of dark matter, we get the constraint
 \begin{equation}
 M \lesssim 5 \times 10^{14} \,    \sqrt{{\rm GeV} \over T_R}      \;{\rm GeV} \;,
 \end{equation}
 which is weaker than the corresponding radiation domination bound.
 Combining it with the constraint $M>H_R$, one finds       that $T_R$ must be below $10^{13}$ GeV for the above analysis to apply.  
 The limiting value of $T_R$ 
  yields  (\ref{Mbound}).
 
We find therefore  that, in the matter dominated case, the bounds on the abundance and the fermion mass are weaker than those in the radiation domination scenario. 
 
 Our conclusion is that sterile neutrino  production during inflation is insignificant unless it is very heavy, $M \sim 10^8\,$GeV. We next consider fermion production is the postinflationary era.

\section{Fermion production in the  inflaton oscillation epoch}

 After inflation completes, the inflaton field starts oscillating around its local minimum. This creates a classical time-dependent background which naturally leads to particle production.
 Such particle production takes place even in the absence of direct renormalizable couplings between the inflaton and other fields. Indeed, gravity, both classical and quantum, induces 
 gauge invariant
 Planck--suppressed operators among various fields. Since the inflaton field value as well as the energy scale of the system  after inflation is below the Planck scale, one may use the effective field theory
  (EFT) approach and 
 expand the Lagrangian in terms of operators of increasing dimension. 
 The leading Planck-suppressed operator  containing the inflaton $\phi$ and the fermion $\Psi$ has dimension 5,
 \begin{equation}
{{\cal C}\over M_{\rm Pl}} \, \phi^2 \, \bar \Psi \Psi  \;,
\end{equation}
where ${\cal C}$ is a dimensionless Wilson coefficient.
 To be conservative, here we assume approximate $\phi \rightarrow -\phi$ symmetry such that operators with odd powers of $\phi$, e.g. $\phi  \bar \Psi \Psi$, can be omitted.
 We also assume conserved parity  which forbids the coupling $ \phi^2 \bar \Psi \gamma_5 \Psi$, although this would not bring additional non-trivial effects.
 
 The value of ${\cal C}$ is a free parameter in the EFT description and can only be computed given a complete quantum
gravity theory. In particular, it can be calculated in string theory via an $n$-point function, where higher dimensional operators play an important role in phenomenology \cite{Buchmuller:2005jr}. 
Generally, such couplings have a very different structure compared to those generated by graviton exchange \cite{Garny:2015sjg,Co:2022bgh}, e.g. they are not related to the energy-momentum tensor nor to 
lower order couplings such as $\phi \bar \Psi \Psi$.

 It is important to note that the above operator is not conformally invariant. Quantum gravity effects generally break conformal invariance as manifested by the existence of the Planck scale itself.
 Classical gravity also breaks this symmetry, while  the breaking is proportional 
  to the fermion mass $M$.  In particular, the above operator is induced classically by the oscillating scale factor after inflation \cite{Ema:2015dka,Ema:2016hlw} with ${\cal C} \propto M$.
  However, at the quantum gravity level, there is no relation between    $M$ and ${\cal C}$.  
    
  The operator at hand cannot be suppressed by requiring inflaton {\it shift invariance}, which is often  invoked during inflation.   At small field values around the minimum of the inflaton potential, this symmetry is 
  completely broken. We find  no general arguments which would lead to natural suppression of ${\cal C}$.
  In what follows, we will treat ${\cal C}$ as a free parameter bounded roughly by one, in order for the EFT description to apply.

 At dimension 6, there is an  additional operator
 \begin{equation}
  { C_6 \over M_{\rm Pl}^2} \,  \, \phi^2 \;\bar \Psi  i \skew8 \not \nabla \Psi    ,
\end{equation}
which reduces to the above operator on-shell,  $ i\skew8 \not \nabla \Psi = M \Psi$, such that 
  ${\cal C} =  C_6 \, {M \over M_{\rm Pl}} $. Therefore, we will not consider it separately. Finally, an operator of the form $   \phi \partial_\mu \phi \, \bar \Psi    \gamma^\mu \Psi$ does not bring anything new
  since it is a total derivative as long as the vector current is conserved.

 \subsection{Fermion production rate}
 
 During preheating, we can 
expand the oscillating inflaton field  as
\begin{equation}
\phi^2(t) = \sum_{n=-\infty}^\infty \zeta_n e^{-in \omega t} \;,
\label{expansion}
 \end{equation}
   where $\omega$ is the oscillation frequency and 
  the coefficients $\zeta_n$ are $slow$  functions of time. 
 An oscillating background  generally entails particle production.

The amplitude  ${\cal M}$   for the {\it Dirac} fermion-antifermion pair production from the ``vacuum''  due to the dim-5 operator is
 \begin{equation}
-i \int_{-\infty}^\infty dt \langle f | V(t) | i \rangle = - i \,{ {\cal C}\over M_{\rm Pl} } \, (2\pi)^4 \delta ({\bf{p}} + {\bf{q}}) \sum_{n=1}^\infty \zeta_n \,\delta(E_p +E_q -n \omega) \;  \bar u v \;,
 \end{equation}
 where $V(t)$ is the interaction term; $p,q$ are the 4-momenta of the created particles and 
 $u,v$ are the relevant Dirac spinors with  given momenta.
 For a fixed $n$ and neglecting the final state particle masses, we get
 \begin{equation}
 \sum_{\rm spin} |{\cal M}_n |^2 = { {\cal C}^2\over M_{\rm Pl}^2 } \, 2 (n\,\omega)^2 \; |\zeta_n|^2 \;.
 \end{equation}
The reaction rate per unit volume is obtained by integrating over the phase space $\Pi$ and summing the contributions for different $n$,  
  \begin{equation}
 \Gamma = \sum_{n=1}^\infty \Gamma_n =   \sum_{n=1}^\infty   \int  \left( \sum_{\rm spin} |{\cal M}_n|^2 \right) d \Pi  =
 {{\cal C}^2 \over 4 \pi   M_{\rm Pl}^2 }        \omega^2  \sum_{n=1}^\infty n^2\,  |\zeta_n|^2   \;.
 \label{Gamma-ss}
  \end{equation}

   In the $Majorana$ fermion case, there are two 
  identical particles in the final state. Hence, the correct amplitude can be obtained with 2 different contractions, 
  ${\cal M}_{\rm Maj} = 2 {\cal M}_{}$. On the other hand, the phase space integral receives
 the factor of 1/2 due to the identical particles. Thus,
  \begin{equation}
 \Gamma_{\rm Maj} = 2  \Gamma_{}  =  {{\cal C}^2 \over 2 \pi   M_{\rm Pl}^2 }        \omega^2  \sum_{n=1}^\infty n^2\,  |\zeta_n|^2 \;.
  \end{equation}

 \subsection{Relic abundance}
 
Unless ${\cal C}$ is large,  particle production proceeds in a rather mild manner such that backreaction of the produced fermions can be neglected.
Then, the particle density for  Dirac fermions
  is found via the Boltzmann equation,
 \begin{equation}
 \dot n + 3Hn = 2\Gamma \;,
 \end{equation}
  where 
  the dot denotes differentiation with respect to coordinate time  $t$ ($ds^2 = dt^2 - a(t)^2 d {\bf x}^2$) and 
  the factor of 2  comes from two particles produced in each reaction.
 The LHS can be written as $a^{-3} {d\over dt } (n a^3)$. 
 To compute the integral, it is convenient to switch to variable $a$, $dt = da /(aH)$, such that
  \begin{equation}
n\, a^3 = \int da \, a^2 \, {2\Gamma \over H} \;.
 \end{equation}

 At this stage, we need to choose the $local$ inflaton potential. Let us start with the quadratic potential and   generalize the result to the quartic potential later. We take
 \begin{equation}
   V= {1\over 2} m_\phi^2 \phi^2    ~~,~~ \phi  (t) = {\phi_0 \over a^{3/2}} \, \cos m_\phi t \;,  
   \end{equation}
  where $\phi_0$ is the initial inflaton amplitude at $a=1$. 
 In this case,  $\omega =    2 m_\phi \, , \, \zeta_1 = {1\over 4}   {\phi_0^2 \over a^3} \;. $
 Thus, for $Dirac$ fermions,
\begin{equation}
\Gamma = {{\cal C}^2 m_\phi^2 \over 16 \pi M_{\rm Pl}^2 } \; {\phi_0^4 \over a^6} \;,
\end{equation}
while Hubble rate is 
\begin{equation}
H = {m_\phi \phi_0 \over {\sqrt{6}}      M_{\rm Pl} \, a^{3/2}} \;.
\label{He}
\end{equation}
The integral is dominated by the initial moments after inflation and soon thereafter the number density becomes
 \begin{equation}
 n(t) = {{\cal C}^2 m_\phi \over 2\sqrt{6} \pi M_{\rm Pl} }  {\phi_0^3 \over a^3} \;,
  \end{equation}
meaning that the  total particle number is conserved. 

The relic abundance generally depends on how quickly reheating occurs.
The matter-dominated (or non-relativistic)  expansion period is characterized by 
\begin{equation}
\Delta_{\rm NR} = \left( {  H_{e} \over H_R }\right)^{1/2} = a_R^{3/4} \;,
\end{equation} 
 where $H_e$ and $H_R$ are the Hubble rates at the end of inflation and reheating, respectively. Here we have defined $a_e=1$ at the end of inflation.
 $\Delta_{\rm NR} \geq 1$ represents the ``dilution'' factor:
 since the produced fermions  are relativistic, 
 the  matter-dominated expansion period dilutes their energy density. It can also be written as the ratio of the reheating temperature in case of instant reheating over the actual reheating temperature,
  $\Delta_{\rm NR}  \simeq T_R^{\rm inst}/ T_R$.
 
Solving for $T_R$, we find the relic abundance
  \begin{equation}
  Y = 10^{-1} \; {\cal C}^2 \;{  H_{e}^{3/2} \, M_{\rm Pl}^{1/2} \over \Delta_{\rm NR}^{} \, m_\phi^2 } 
  \end{equation}
  for $g_* =107$, which makes the diluting effect of the matter-dominating era explicit. 

\subsection{Constraints and implications}

It is more convenient to trade 
$m_\phi$ for $\phi_0$, which allows us to derive a universal result which is also valid for the quartic inflaton potential. 
 Requiring $Y$ not to exceed that of dark matter, we get
 \begin{equation}
 {\cal C} \lesssim 10^{-4} \; \Delta_{\rm NR}^{1/2}\;  { {M_{\rm Pl}^{3/4} }  H_e^{1/4}      \over \phi_0}  \; \sqrt{{{\rm GeV} \over M }} \;,
 \label{constr-C}
 \end{equation}
 as long the fermion effective mass is below the inflaton mass.
This result applies to the quartic inflaton potential   $V={1\over 4} \lambda_\phi \phi^4$ as well.
In this case, 
 $\omega \simeq 0.85 \sqrt{\lambda_\phi} \phi_0/a$ and $\zeta_1 \simeq 0.25 \phi_0^2/a^2$, leading to 
 $Y \simeq 1.2 \times 10^{-2} \; {\cal C}^2 \; {\phi_0^2 \over {M_{\rm Pl}^{3/2} }  H_e^{1/2}}$. Then, the above constraint applies irrespective of the 
 reheating temperature and 
 \begin{equation}
 \Delta_{\rm NR} \simeq 1 
 \end{equation}
in the quartic case. The Dirac and Majorana fermion bounds  on ${\cal C}$ are very similar and only differ by a square root of 2. 

 To understand the strength of the above  bound, let us take the typical large-field inflation values   $\phi_0 \sim M_{\rm Pl}$ and 
 $H_{e}\sim 10^{-5} M_{\rm Pl}$.
  Then,
  \begin{equation}
 {\cal C} \leq  10^{-5}  \; \Delta_{\rm NR}^{1/2} \; \sqrt{{{\rm GeV} \over M }} \;.
  \end{equation}
 Therefore, unless the dilution factor is very large, the Wilson coefficient has to be very small, ${\cal C} \ll 1$, for GeV or above GeV scale fermion masses.
 Otherwise, the Universe would be too dark.  The constraint on ${\cal C}$ is weaker than the corresponding bound on the Wilson coefficient of the dim-6 operator $\phi^4 s^2/M_{\rm Pl}^2$
 for a scalar dark relic  $s$ \cite{Lebedev:2022cic}, as expected.
 
  This result implies that   the constraints on the Wilson coefficients of  higher dimensional operators,
 \begin{equation}
 {1\over M^3_{\rm Pl}} \phi^4 \bar \Psi \Psi   ~,~  {1\over M^5_{\rm Pl}} \phi^6 \bar \Psi \Psi   ~,~...
  \end{equation} 
 are also non-trivial for GeV masses or above. They are weaker roughly by the factor $(M_{\rm Pl} /\phi_0)^p$, where $2p+2$ is the power of the inflaton field in the corresponding operator.  
 As long as $\phi_0$ is not far below the Planck scale, the constrains are significant. Therefore, full control over these operators   is necessary in order to make reliable predictions.

The fermion--inflaton coupling creates an effective fermion mass during inflation, which may suppress inflationary particle production. For example, if one 
    blindly   extrapolates the coupling 
$  {\cal C}    \phi^2 \bar \Psi \Psi  /M_{\rm Pl}$ to large inflaton field values (where  the expansion in $\phi/M_{\rm Pl}$ breaks down),
 the effective mass would be of order ${\cal C}  \phi^2/M_{\rm Pl}$.
  This could be larger than the inflationary Hubble rate $H$, in which case particle production via inflation would be suppressed. This issue is model-dependent and since inflationary particle production is not the leading effect in any case, 
it is insignificant   for our purposes.

An important conclusion we make from the above bound is that 
  the quantum-gravity generated  operator $\phi^2 \bar \Psi \Psi  /M_{\rm Pl}$ with a small Wilson coefficient can generate {\it all of the dark matter}.
  In particular, keV scale sterile neutrinos can play the role of dark matter for 
  \begin{equation}
  {\cal C} (M\sim {\rm keV}) \simeq 10^{-2} - 10^{-1} \;,
  \label{range}
  \end{equation} 
 when $ \Delta_{\rm NR} \sim 1 $.  This follows from (\ref{constr-C}) with $H_e \sim 10^{-5}\,M_{\rm Pl}$ and $\phi_0 \leq 0.1 \, M_{\rm Pl}$. Such an initial inflaton value is required to avoid kinematic suppression of the inflaton decay. Indeed, the effective neutrino mass during the inflaton oscillation epoch is $  {\cal C} \phi^2$, which should be below the inflaton mass for efficient particle production. Assuming a $locally$ quadratic inflaton potential,  the inflaton mass is constrained by the Hubble rate at the end of inflation, $m_\phi  =  \sqrt{6} H_e M_{\rm Pl} /\phi_0$. Combining this with the kinematic constraint, one
 obtains   $\phi_0 \leq 0.1 \, M_{\rm Pl}$ and the   ${\cal C}$-range (\ref{range}) when the sterile neutrino mass is varied between 1 and 5 keV.

 Such neutrinos, and even much lighter ones,  would constitute {\it cold } dark matter. This can be seen as follows \cite{Lebedev:2022vwf}: their initial energy $E_\Psi$ is of order $m_\phi^{\rm eff}$, which is $m_\phi$ for the quadratic and $\sqrt{\lambda_\phi} \phi_0$ for the quartic inflaton potential. On the other hand, the SM bath temperature $T$ is determined by the scale of the inflaton potential $V^{1/4} (\phi)$. Since
 \begin{equation}
 E_\Psi \sim m_\phi^{\rm eff} \ll V^{1/4} \sim T \;
 \end{equation} 
 as long as $m_\phi \ll \phi_0$ and $\lambda_\phi \ll 1$, the typical neutrino energy in the relativistic regime  is far below the SM bath temperature. In the quadratic case, there is a further suppression factor due to the redshifting
 of the relativistic neutrino energy relative to $ V^{1/4}$. As a result, sterile neutrinos  become non-relativistic at $T \gg M$ and thus are ``cold'' at the stage of structure formation. In contrast, $\nu_R$ produced via the Dodelson-Widrow mechanism have the energy related to the temperature of the thermal bath and hence are ``warm'', which makes this scenario disfavored.  
 
 Decaying sterile neutrinos can produce a range of signatures, in particular, monochromatic photons. The latter could fall in the X-ray range as in the Dodelson-Widrow model, although
 this possibility is not theoretically favored over the others. Heavier decaying neutrinos would lead to diffuse gamma ray emission, which provides us with another avenue to probe
 sterile neutrino dark matter.

  \section{Conclusion}
  
 We have studied production of feebly interacting fermions, in particular, sterile neutrinos, via gravitational effects during and immediately after  high-scale   inflation. 
 We find that these effects are important and lead to a background of long-lived or stable relics, which can account for all of the dark matter.
 
 Sterile neutrino  production via classical gravity during inflation  is suppressed by its mass $M$, which represents a conformal symmetry breaking parameter.
If the Universe is dominated by radiation after inflation, the neutrino abundance is given  by $Y \sim 10^{-2} \,(M/M_{\rm Pl})^{3/2}$
irrespectively of the inflationary Hubble rate $H \gg M$ and the reheating temperature $T_R$. In the case of extended matter domination, the abundance becomes 
$Y \sim 10^{-2} \,M \,T_R/M_{\rm Pl}^2$.
In either case, the sterile neutrino abundance is negligible unless $M \gtrsim 10^8\,$GeV.

After inflation, particles are efficiently produced due to inflaton  ($\phi$) oscillations. In order to account  for quantum gravity effects, we resort to the EFT description at $\phi < M_{\rm Pl}$ and
expand the inflaton-neutrino interactions in inverse powers of $M_{\rm Pl}$, focussing on the lowest order 
  Planck-suppressed operators such as  $\phi^2 \bar \Psi \Psi /M_{\rm Pl}$.  
  Since quantum gravity breaks conformal invariance, the Wilson coefficients of these operators are not related to the mass parameter $M$ and we treat them as free parameters. 
  We find that the above operator and its higher dimensional analogs are very efficient in particle production. Even if 
   the corresponding Wilson coefficient   is very small,  $\phi^2 \bar \Psi \Psi /M_{\rm Pl}$ can readily produce all of the Universe dark matter in the form of long-lived sterile neutrinos.
   This is the case even for keV scale  (or below)  sterile neutrinos. The energy of these neutrinos is not related to the temperature of the SM thermal bath $T$, unlike it is in the Dodelson-Widrow mechanism.
   They become non-relativistic at $T \gg M$ and thus constitute $cold$ dark matter, which is favored by the structure formation constraints.
   
   The gravitational production mechanism is operative irrespective of the active-sterile mixing angle.
It  creates a background for other sterile neutrino production models such as  freeze-in, etc. \cite{Petraki:2007gq,Merle:2013wta,Adulpravitchai:2014xna,Drewes:2015eoa,Bringmann:2022aim}.
  If the mixing angle is not too small, one expects to see signatures of decaying dark matter, for instance,  monochromatic photons.
  This possibility remains viable in the X-ray range, as was initially expected in the Dodelson-Widrow model.   
   \\ \ \\
   \noindent 
   {\bf Acknowledgements.} 
   OL acknowledges support by Institut Pascal at Universit\'e Paris-Saclay during the Paris-Saclay Astroparticle Symposium 2023, with the support of the P2IO Laboratory of Excellence (program ``Investissements d'aveni'' ANR-11-IDEX-0003-01 Paris-Saclay and ANR-10-LABX-0038), the P2I axis of the Graduate School of Physics of Universit\'e Paris-Saclay, as well as IJCLab, CEA, IAS, OSUPS, and the IN2P3 master project UCMN. The research of S.P. has received partial financial support by the National Science Centre, Poland, grant DEC-2019/35/B/ST2/02008.

\end{document}